\documentclass[%
 reprint,
 amsmath,
 amssymb,
 aps,
 pra,
 showpacs
]{revtex4-1}

\usepackage{graphicx}
\usepackage{dcolumn}
\usepackage{bm}
\usepackage{bbold}
\usepackage[export]{adjustbox}
\usepackage{amsmath,amsfonts,mathtools}
\usepackage{dsfont}
\usepackage{amsmath}
\usepackage{amssymb}
\usepackage{makeidx}
\usepackage{graphicx}
\usepackage{color}

\newcommand{\wrt}{w.r.t.\ }

\newcommand{\hide}[1]{}

\newcommand{\im}{\Im m\,}

\newcommand{\om}{\omega}

\newcommand{\lcase}{\left\{\begin{array}{ll}}
\newcommand{\rcase}{\end{array}\right.}
\newcommand{\ear}{\end{array}}
\newcommand{\bal}{\begin{align}}
\newcommand{\eal}{\end{align}}
\newcommand{\bma}{\begin{pmatrix}}
\newcommand{\ema}{\end{pmatrix}}
\newcommand{\beq}{\begin{equation}}
\newcommand{\eeq}{\end{equation}}
\newcommand{\bel}[1]{\begin{equation}\label{eq:#1}}
\newcommand{\eel}{\end{equation}}
\newcommand{\bea}{\begin{eqnarray}}
\newcommand{\eea}{\end{eqnarray}}
\newcommand{\beaNN}{\begin{eqnarray*}}
\newcommand{\eeaNN}{\end{eqnarray*}}

\newcounter{lecture}

\newcommand{\Ef}{\mathcal{E}}

\newcommand{\pde}{\partial}
\renewcommand{\hide}[1]{}

\newcommand{\rev}[1]{{ #1}}
\newcommand{\inten}[2]{#1\times10^{#2}\,\text{W/cm}^2}
\newcommand{\optc}{\text{ opt.cyc.}}
\renewcommand{\im}{\text{i}}

\newcommand{\hydroplus}{\text{H}_2^+}
\newcommand{\hydrogen}{\text{H}_2}

\begin{document}

\preprint{APS/123-QED}

\title{Theoretical investigation of Freeman resonance in the dissociative ionization of $H_2^+$}
\author{Jinzhen Zhu}
\email{Jinzhen.Zhu@physik.uni-muenchen.de,zhujinzhenlmu@gmail.com}
\affiliation{%
 Physics Department, Ludwig Maximilians Universit\"at, D-80333 Munich, Germany
}%
\begin{abstract}
The dissociative ionization of $H_2^+$ in linearly polarized, 400 nm laser pulses is simulated by solving a three-particle time-dependent Schr\"odinger equation in full dimensionality.
The joint energy spectra (JES) are computed for $\cos^8$ and flat-top envelopes using the time-dependent surface flux (tSurff) methods.
In JES, the energy sharing of $N$ photons with frequency $\omega$ by nuclear kinetic energy release (KER) $E_N$ and electronic KER $E_e$ is well described by $E_N+E_e=N\omega-U_p+E_0$ for $\cos^8$ envelope, but satisfy $E_N+E_e=N\omega+E_0$ for flat-top envelope, exposing a deviation of the ponderomotive energy $U_p$, where $E_0$ is the ground energy of $H_2^+$ and this observation has been observed in experiments.
The analysis of the wavefunction for electrons and protons after the pulse are presented, where we find $U_p$ is absorbed by the Freeman resonances between two excited ungerade states of $H_2^+$.
\end{abstract}

\pacs{32.80.-t,32.80.Rm,32.80.Fb}
\maketitle

\section{\label{sec:intro}Introduction}
Being a typical candidate for the investigation of the tree-body Coulomb interaction problem in attosecond physics, $\hydroplus$ has been investigated a lot both in experimental and theoretical sides~\cite{Bucksbaum1990,Zuo1995,Yao1993,Giusti-Suzor1995,Jolicard1992,Zuo1993,Posthumus2004,Giusti-Suzor1990,Yue2013,Yue2014,Odenweller2011,Odenweller2014,Wu2013,Gong2016}.
The JES of the KER for one electron $E_e$ and two protons $E_N$ of the $\hydroplus$ ion are predominant observables that show how energy is distributed around the fragments.
\par
As required by quantum mechanics and law of energy conservation, the equation $E_N+E_e=N\omega+E_0-U_p$ for multi-photon ionization is preferred in JES, which means that the total energy of the three particles is $N$ photons $\omega$ subtracted by the ponderomotive energy of the electron $U_p$.
Wu et al.~\cite{Wu2013} reported the energy sharing of fragments of $\hydrogen$ in the JES in experiments using 400 nm, linearly polarized and long pulses.
Their observations are also helpful for the investigation of $\hydroplus$, because in experiments, $\hydroplus$ are created from $\hydrogen$.
In their experiments $E_N+E_e$ lines do not move considerably for intensities $\inten{5.9}{13}$ and $\inten{4.3}{13}$, which means the contribution of intensity dependent $U_p$ is missing, because of Freeman resonances~\cite{Freeman1987}.
However, relative theoretical studies on a 400 nm computation have not been reported yet. 
\par
The difficulties for theoretical studies mainly comes from the computational cost that grows exponentially with the wavelength and intensities.
The scaling problem can be relived by using the time-dependent surface flux method (tSurff)~\cite{Tao2012}, which has been applied to another three-body system, He, in full dimensionality~\cite{Zielinski2016, Zhu2020}, and also has been successfully applied to 2D models~\cite{Yue2013,Yue2014}.
The dissociative ionization of the $\hydroplus$ has also been simulated in reduced dimensionality by other groups~\cite{Steeg2003,Qu2002,Silva2013,Madsen2012,Odenweller2011,Takemoto2010,Feuerstein2003,Kulander1996}, where the nuclear KER
is most probable around 0.5 atomic units, far from experimental observations.
Our previous paper~\cite{Zhu2020b} reported the dissociative ionization of $\hydroplus$ ion in full dimensionality and gives the nuclear KER (2$\sim$4 eV) close to experimental observables~\cite{Odenweller2014,Wu2013,Gong2016}.
\par
In this paper we will investigate the Freeman resonance of dissociative of $\hydroplus$ by quantum simulations in full dimensionality based on the tRecX code.
\section{Computational Details}
In this paper, atomic units $\hbar=e^2=m_e=4\pi\epsilon_0\equiv1$ are used if not specified.
Spherical coordinates with center of the mass of two protons as the origin are applied.
Instead of using the vector between two protons $\vec{R}$ as coordinate~\cite{Yue2013,Yue2014,Madsen2012}, we specify the coordinates of the protons and electrons as $\vec{r_1},-\vec{r_1}$ and $\vec{r_2}$.
We denote $M=1836$ atomic units as the mass of the proton.
\subsection{Hamiltonian}
The wavefunction is depicted by $\psi(\vec{r_1},\vec{r_2},t)$ that satisfies $\im\partial_t \psi(\vec{r_1},\vec{r_2},t)= H\psi(\vec{r_1},\vec{r_2},t)$.
The total Hamiltonian can be represented by the sum of the electron-proton interaction $H_{EP}$ and two tensor products, written as
\begin{equation}\label{eq:HamiltonianH2PlusFull}
 H=H^{(+)}\otimes \mathds{1}+\mathds{1}\otimes H^{(-)}+H_{EP},
\end{equation}
where the tensor products are formed by the identity operator $\mathds{1}$ multiplied by the Hamiltonian for two protons ($H^{(+)}$) or for the electron ($H^{(-)}$).
With the coordinate transformation used in Ref.~\cite{Hiskes1961}, the single operator for the electron is
\begin{equation}
 H^{(-)}=-\frac{\Delta}{2m}-\im\beta\vec{A}(t)\cdot\vec{\triangledown },
\end{equation}
and the Hamiltonian for protons can be written as
\begin{equation}
  H^{(+)}=-\frac{\Delta}{4M}+\frac{1}{2r},
\end{equation}
where we introduce reduced mass $m=\frac{2M}{2M+1}\approx1$ and $\beta=\frac{1+M}{M}\approx1$ for the electron.
The Hamiltonian of the electron-proton interaction can be written as
\begin{equation}
 H_{EP}=-\frac{1}{|\vec{r_1}+\vec{r_2}|}-\frac{1}{|\vec{r_1}-\vec{r_2}|}.
\end{equation}

\rev{
\subsection{Computational methods}
The tSurff method is applied for computing the JES and is detailed in our previous works~\cite{Zhu2020b,Zielinski2016}.
A simplified version with important points is presented here for completion.\par
Following the essence of the tSurff method, we neglect the interactions of protons and electrons beyond a sufficient large tSurff radius $R_c^{(+/-)}$, with the corresponding Hamiltonians being $H_V^{(+)}=-\frac{\Delta}{4M}$ for two protons and $H_V^{(-)}=-\frac{\Delta}{2m}-\im\beta\vec{A}(t)\cdot\vec{\triangledown}$ for the electron.
The scattered states of the two protons and the electron are Volkov solutions, which satisfy $\im\partial_t \chi_{\vec{k}_1}(\vec{r}_1)=H_V^{(+)}\chi_{\vec{k}_1}(\vec{r}_1)$ and $\im\partial_t \chi_{\vec{k}_2}(\vec{r}_2)=H_V^{(-)}\chi_{\vec{k}_2}(\vec{r}_2)$, respectively,
where $\vec{k}_{1/2}$ denote the momenta of the protons or the electron.
\par
Based on the tSurff radius $R_c^{(+/-)}$, we may split the dissociative ionization into four regions namely $B,I,D,DI$, shown in Fig.~\ref{fig:H2PlusRegions}, where bound region $B$ preserves the full Hamiltonian in Eq.~(\ref{eq:HamiltonianH2PlusFull}), $D,I$ are time propagations by single particles with the Hamiltonians being
\begin{equation}
 H_{D}(\vec{r}_2,t)=H_{V}^{(-)}(\vec{r}_2,t) = -\frac{\Delta}{2m}-\im\beta\vec{A}(t)\cdot\vec{\triangledown}
\end{equation}
and
\begin{equation}
 H_{I}(\vec{r}_1,t) = -\frac{\Delta}{4M}+\frac{1}{2r_1},
\end{equation}
and $DI$ is an integration process.
The treatment was first introduced in the double ionization of Helium in Ref.~\cite{Scrinzi2012} and then applied in a 2D simulation of the $\hydroplus$ ion in Ref.~\cite{Yue2013}.
\begin{figure}
\centering
\includegraphics[width=0.4\textwidth]{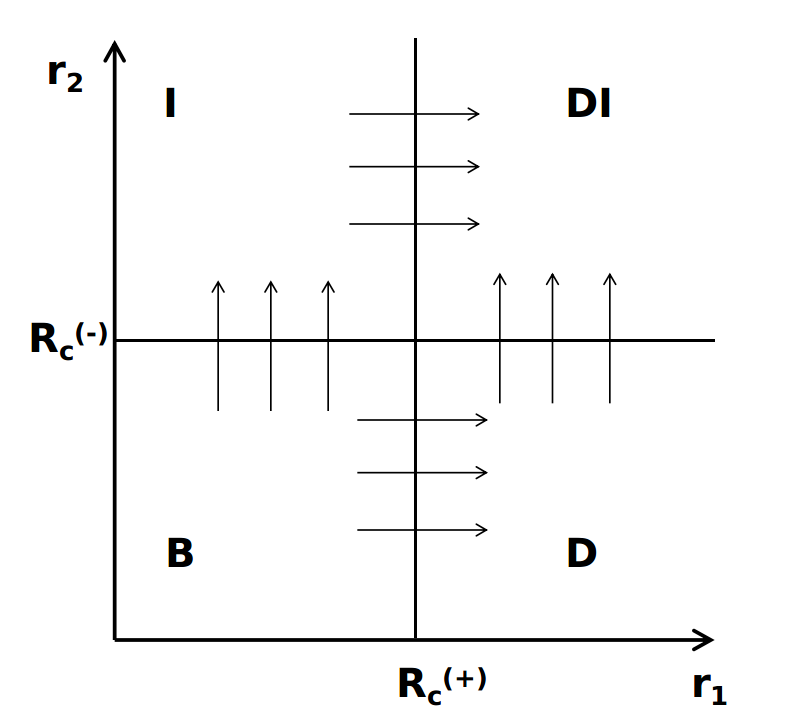}
\caption{The regions of dissociative ionization time propagation.
The B stands for bound region, D for dissociation region where the two protons are out of $R_c^{(+)}$ but electron not ionized and stays inside.
I represents the ionization region where electron is out-of-box $R_c^{(-)}$ but two protons are still inside $R_c^{(+)}$.
DI stands for the dissociative ionization region where both the electron and the protons are out of $R_c^{(+/-)}$. $R_c^{(+/-)}$ are the tSurff radii for $r_1=|\vec{r}_1|$ or $r_2=|\vec{r}_2|$.
}
\label{fig:H2PlusRegions}
\end{figure}
\par
We assume that for a sufficiently long propagation time $T$, the scattering ansatz of the electron and protons disentangle.
By introducing the step function 
\begin{equation}
 \Theta_{1/2}(R_c)=\left\{\begin{matrix}
0\;, r_{1/2}< R_c^{(+/-)} \\ 
1\;, r_{1/2}\geq  R_c^{(+/-)},
\end{matrix}\right.
\end{equation}
the unbound spectra can be written as
\begin{equation}\label{eq:finalSpectrum}
 P(\vec{k}_1,\vec{k}_2)=\left | b(\vec{k}_1,\vec{k}_2,T) \right |^2,
\end{equation}
where the scattering amplitudes $b(\vec{k}_1,\vec{k}_2,T)$ are
\begin{equation}\label{eq:integralAmplitudes}
\begin{split}
b(\vec{k}_1,\vec{k}_2,T)=&\langle \chi_{\vec{k}_1}\otimes \chi_{\vec{k}_2} |\Theta_1(R_c)\Theta_2(R_c)|\psi(\vec{r}_1,\vec{r}_2,t)\rangle \\
=&\int_{-\infty}^{T}[F(\vec{k}_1,\vec{k}_2,t)+\bar{F}(\vec{k}_1,\vec{k}_2,t)] dt
\end{split}
\end{equation}
with two sources being
\begin{equation}\label{eq:Fk1k2H2Plus}
 F(\vec{k}_1,\vec{k}_2,t)= \langle \chi_{\vec{k}_2}(\vec{r}_2,t)\left |[H_V^{(-)}(\vec{r}_2,t), \Theta_2(R_c)] \right | \varphi_{\vec{k}_1}(\vec{r}_2,t) \rangle
\end{equation}
and 
\begin{equation}\label{eq:Fk1k2H2PlusBar}
 \bar{F}(\vec{k}_1,\vec{k}_2,t)= \langle \chi_{\vec{k}_1}(\vec{r}_1,t)\left | [H_V^{(+)}(\vec{r}_1,t), \Theta_1(R_c)] \right | \varphi_{\vec{k}_2}(\vec{r}_1,t) \rangle.
\end{equation}
The single particle wavefunctions $\varphi_{\vec{k}_1}(\vec{r}_2,t)$ and $\varphi_{\vec{k}_2}(\vec{r}_1,t)$ satisfy
\begin{equation}\label{eq:unbound0}
 \im\frac{d}{dt}\varphi_{\vec{k}_1 }(\vec{r}_2,t)=H_D(\vec{r}_2,t) \varphi_{\vec{k}_1 }(\vec{r}_2,t)-C_{\vec{k}_1 }(\vec{r}_2,t)
\end{equation}
and
\begin{equation}\label{eq:unbound1}
 \im\frac{d}{dt}\varphi_{\vec{k}_2 }(\vec{r}_1,t)=H_I(\vec{r}_1,t)\varphi_{\vec{k}_2 }(\vec{r}_1,t)-C_{\vec{k}_2 }(\vec{r}_1,t).
\end{equation}
The sources are the overlaps of the two-electron wavefunction and the Volkov solutions shown by 
\begin{equation}\label{eq:source1}
 C_{\vec{k}_1 }(\vec{r}_2,t)=\int d\vec{r}_1 \overline{\chi_{\vec{k}_1}(\vec{r}_1,t)}[H_V^{(+)}(\vec{r}_1,t),\Theta_1(R_c)]\psi(\vec{r}_1,\vec{r}_2,t)
\end{equation}
and
\begin{equation}\label{eq:source2}
 C_{\vec{k}_2 }(\vec{r}_1,t)=\int d\vec{r}_2 \overline{\chi_{\vec{k}_2}(\vec{r}_2,t)}[H_V^{(-)}(\vec{r}_2,t),\Theta_2(R_c)]\psi(\vec{r}_1,\vec{r}_2,t),
\end{equation}
with initial values being 0, where $\overline{\cdots}$ means complex conjugate.
The two tSurff radii could be set as equivalent $R_c^{(+)}=R_c^{(-)}$, because all Coulomb interactions are neglected when either the protons or electron is out of the tSurff radius.
According to our previous researches, the spectrum computation is independent of the $R_c$ if all Coulomb terms are removed and the wavefunction is propagated long enough after the pulse~\cite{Zielinski2016,Scrinzi2012,Zhu2020b}.
Apart from the tSurff method, the infinite-range exterior complex scaling (irECS) method is utilized as an absorber~\cite{Scrinzi2010}.
}

\subsection{Laser pulses}
\label{sec:pulses}
The dipole field of a laser pulse with peak intensity $I=\Ef_0^2$ (atomic units, $\Ef_0$ is the peak electric field) and linear polarization in the $z$-direction is defined as $\Ef_z(t)=\pde_tA_z(t)$ with vector potential
\begin{equation}
 A_z(t)=\frac{\Ef_0}{\om} a(t)\sin(\om t+\phi_{CEP}),
\end{equation}
where $\phi_{CEP}$ is the phase of the pulse.
The pulses with wavelength $\lambda=400$ nm are applied with peak intensities $\inten{8.3}{13}$ and $\inten{5.9}{13}$ as used in our previous works~\cite{Zhu2020b}.
Pulse duration of all the pulses are specified as full width at half maximum (FWHM) $T_{FWHM}=5\optc$ \wrt intensity.
We choose $a(t)=[\cos(t/T_{FWHM})]^8$, similar to the Gaussian like envelope to approximate the realistic pulses.   
Apart from the $\cos^8$ envelope, a “flat-top” trapezoidal function with a linear rise and descent over a single optical cycle is also applied to simulate the long pulse.
We use the flat-top envelope pulse because it is suitable to investigate the Freeman resonances.
\section{Numerical results}
The discretization parameters used here are the same as used in our previous work in Ref.~\cite{Zhu2020b}.
The field free ground energy value is $E_0=-0.592$ atomic units and the internuclear distance is 2.05 atomic units.
With the kinetic energy of protons excluded, the ground eigenenergy is -0.597 atomic units, three digits exact to the ground energy from quantum chemistry calculations in Ref.~\cite{Bressanini1997}, where the internuclear distance is fixed.
The internuclear distance is 1.997 atomic units, three digits exact to that from the precise computations in Ref.~\cite{Schaad1970}.
\subsection{Joint energy spectra}
\begin{figure}
\centering
\includegraphics[width=0.23\textwidth,trim=0.1cm 0.1cm 0.1cm 0.1cm,clip]{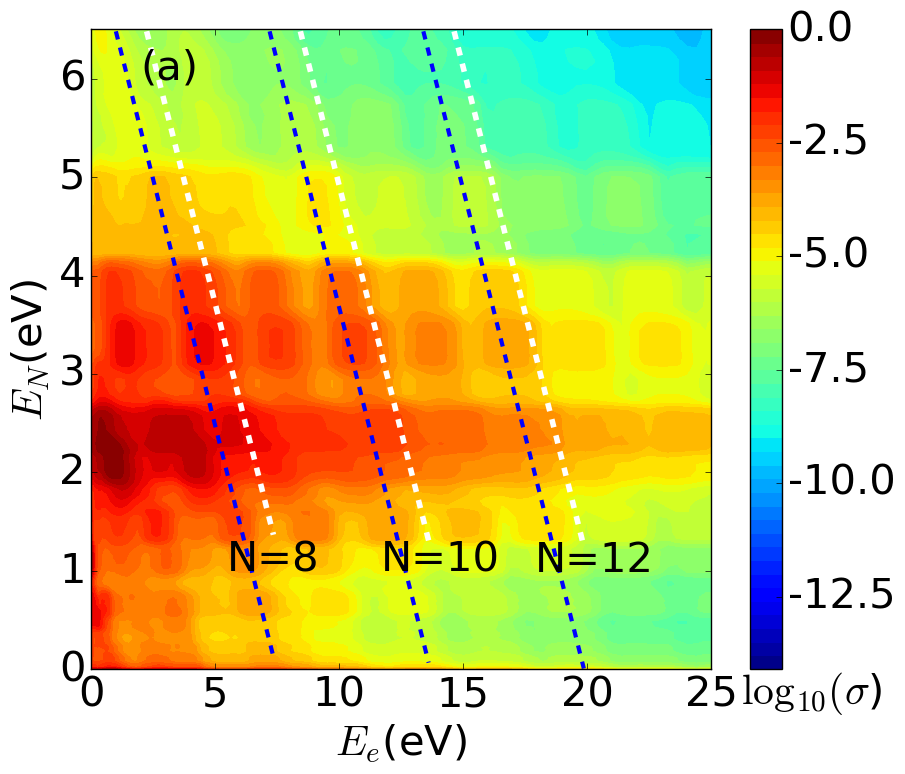}
\includegraphics[width=0.23\textwidth,trim=0.1cm 0.1cm 0.1cm 0.1cm,clip]{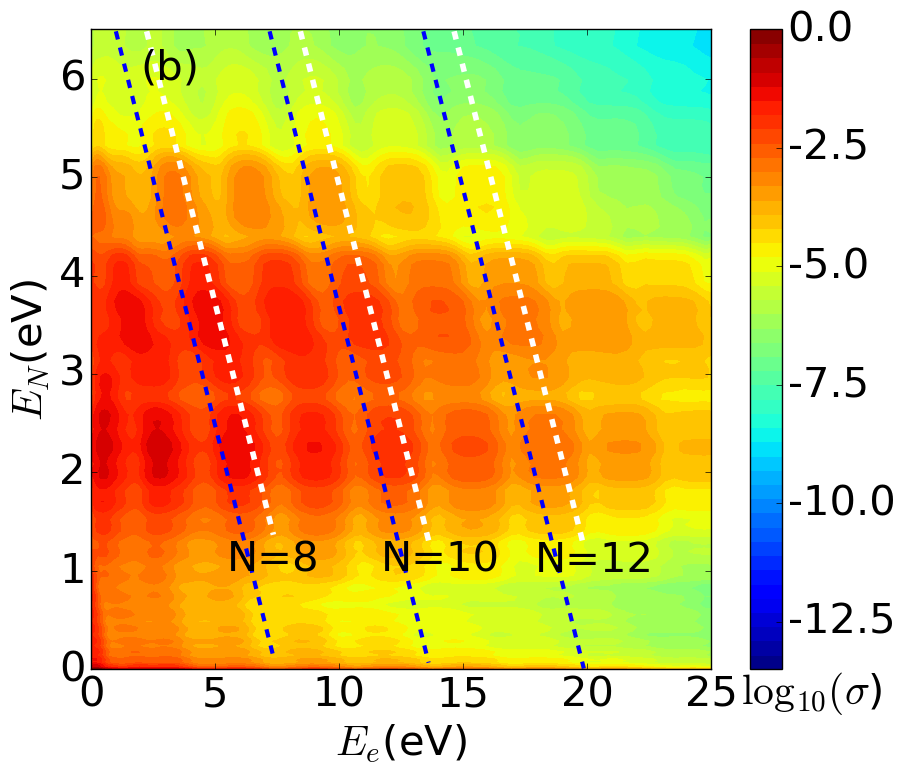}
\includegraphics[width=0.23\textwidth,trim=0.1cm 0.1cm 0.1cm 0.1cm,clip]{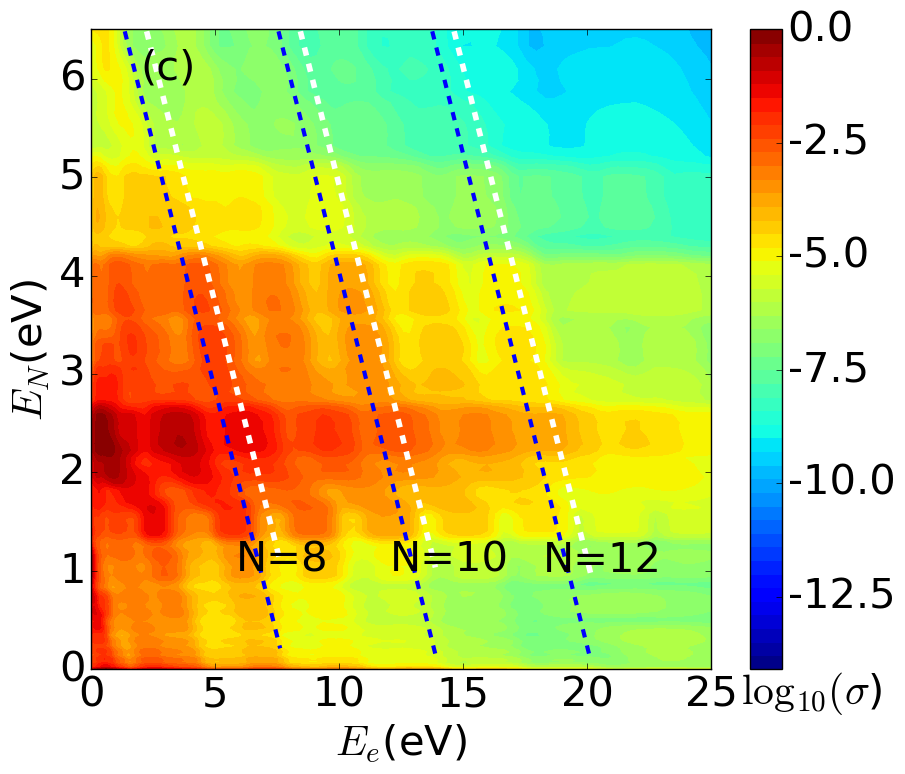}
\includegraphics[width=0.23\textwidth,trim=0.1cm 0.1cm 0.1cm 0.1cm,clip]{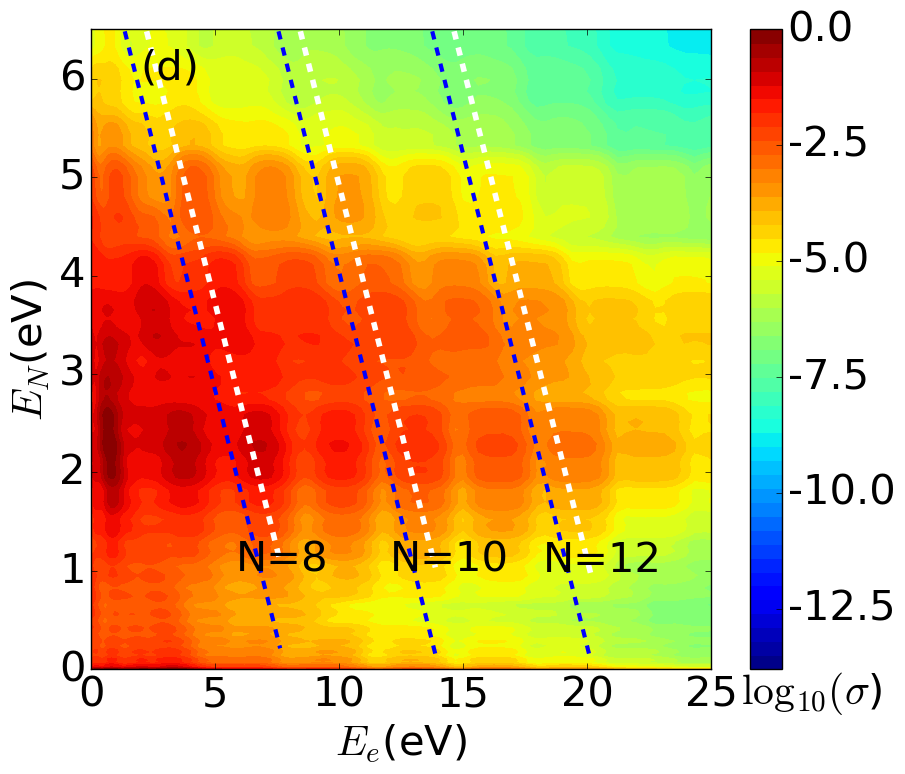}
\caption{Log-scale JES $\log_{10}\sigma(E_{N},E_{e})$ represented by total energy of two protons $E_{N}$ and that of an electron $E_{e}$.
Linear polarized, 400 nm, with (a) $\cos^8$ envelope, $I=\inten{8.3}{13}$, (b) flat-top envelope, $I=\inten{8.3}{13}$, (c) $\cos^8$ envelope, $I=\inten{5.9}{13}$, and (d) flat-top envelope, $I=\inten{5.9}{13}$ with FWHM=5 $\optc$ pulses are applied to the $\hydroplus$ ion. The blue dashed lines represent the energy sharing between the protons and electron with formula $E_{N}+E_{e}=N\omega+E_0-U_p$ and white ones represent $E_{N}+E_{e}=N\omega+E_0$, where $\omega$ is the photon energy.
}
\label{fig:H2PlusJES}
\end{figure}

The JES of the two dissociative protons and the electron is obtained by
\begin{equation}
\begin{split}
 \sigma(E_N,E_e)=&\int d\phi_1\int d\phi_2\int d\theta_1\sin\theta_1\int d\theta_2\sin\theta_2\\
 &P(\phi_1,\theta_1,\sqrt{4M E_N},\phi_2,\theta_2,\sqrt{2m E_{e} }),
 \end{split}
\end{equation}
where $E_N,E_e$ are the KERs of {\it two protons} and an electron, respectively.
The spectrum $P(\vec{k}_1,\vec{k}_2,T)$ is from Eq.~\ref{eq:finalSpectrum}.
$\sigma(E_{N},E_{e})$ is presented in Fig.~\ref{fig:H2PlusJES}.
The blue tilt lines with formula $E_{N}+E_{e}=N\omega+E_0 -U_p$ with $U_p=\frac{A_0^2}{4m}$ specify the energy sharing of $N$ photons for all computations.
For the $\cos^8$ envelope computations in Fig.~\ref{fig:H2PlusJES} (a) and (c), the blue dashed lines representing $E_{N}+E_{e}=N\omega+E_0-U_p$ fall in the peak of the stripe in JES, representing the standard energy sharing signatures.
For the flat-top envelope in Fig.~\ref{fig:H2PlusJES} (b) and (d), although each dashed line $E_{N}+E_{e}=N\omega+E_0-U_p$ labels a certain tilt stripe in JES, it is not at the center of the stripes, which, however, are well described by $E_{N}+E_{e}=N\omega+E_0$.
This behavior was also observed in experiments with long pulses~\cite{Wu2013} and was attributed to Freeman resonances.
One also observes that the JES yield with nuclear KER $\geq 3$ eV are more considerable for flat-top computations, whereas the peak of JES are mainly located with nuclear KER $\leq 3$ eV for $\cos^8$ computations.
\rev{The Freeman resonance is expected with a flat-top envelope laser pulse with constant intensity.}
We will \rev{show that} the "missing of $U_p$" in JES \rev{is from Freeman resonance and investigate the underlying mechanism} in the following section.
\rev{
\par
To verify the above observation with the experimental data from references, a long (FWHM = 13$\optc$), $\cos^8$ shape laser pulse as used in Ref.~\cite{Wu2013} is applied for the calculation, where the JES is shown in Fig.~\ref{fig:H2PlusJESExp} (a).
Because Freeman resonance could be expected with a long pulse~\cite{Freeman1987}.
In Fig.~\ref{fig:H2PlusJESExp} (a), we also find the white lines representing $E_{N}+E_{e}=N\omega+E_0$ are in the middle of the stripes in JES, whereas the blue lines representing $E_{N}+E_{e}=N\omega+E_0-U_p$ are not, indicating a "missing of $U_p$" from Freeman resonance.
Another laser pulse with flat-top envelope is also applied for a comparison with the $\cos^8$ envelope in Fig.~\ref{fig:H2PlusJESExp} (b), where the "missing of $U_p$" is also observed, consistent with the flat-top computation with 400 nm laser pulses above.
\begin{figure}
\centering
\includegraphics[width=0.23\textwidth,trim=0.1cm 0.1cm 0.1cm 0.1cm,clip]{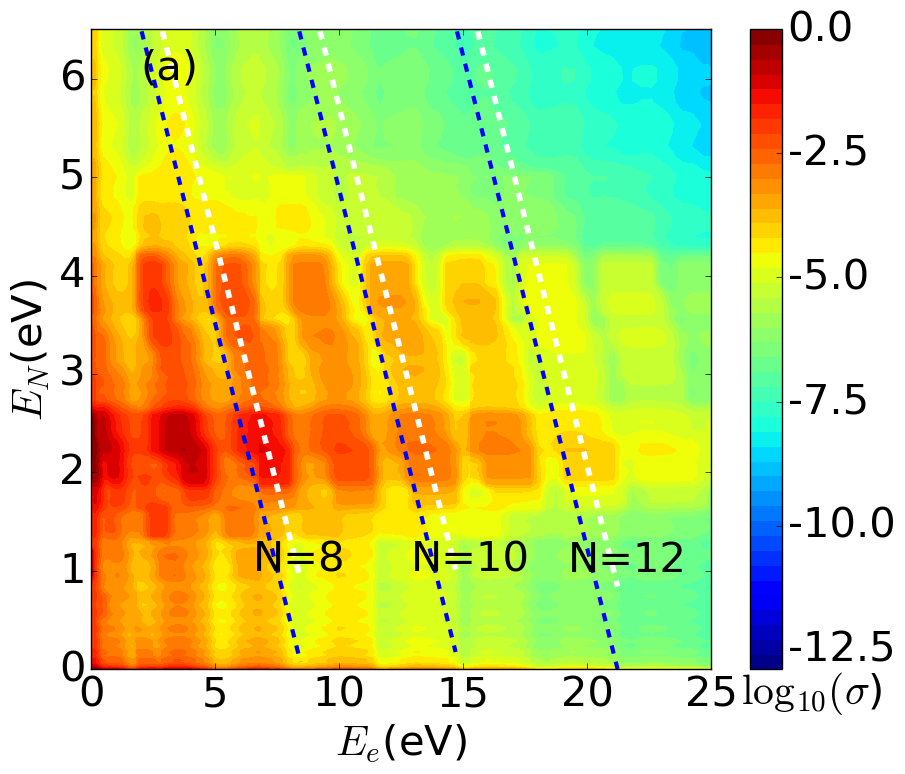}
\includegraphics[width=0.23\textwidth,trim=0.1cm 0.1cm 0.1cm 0.1cm,clip]{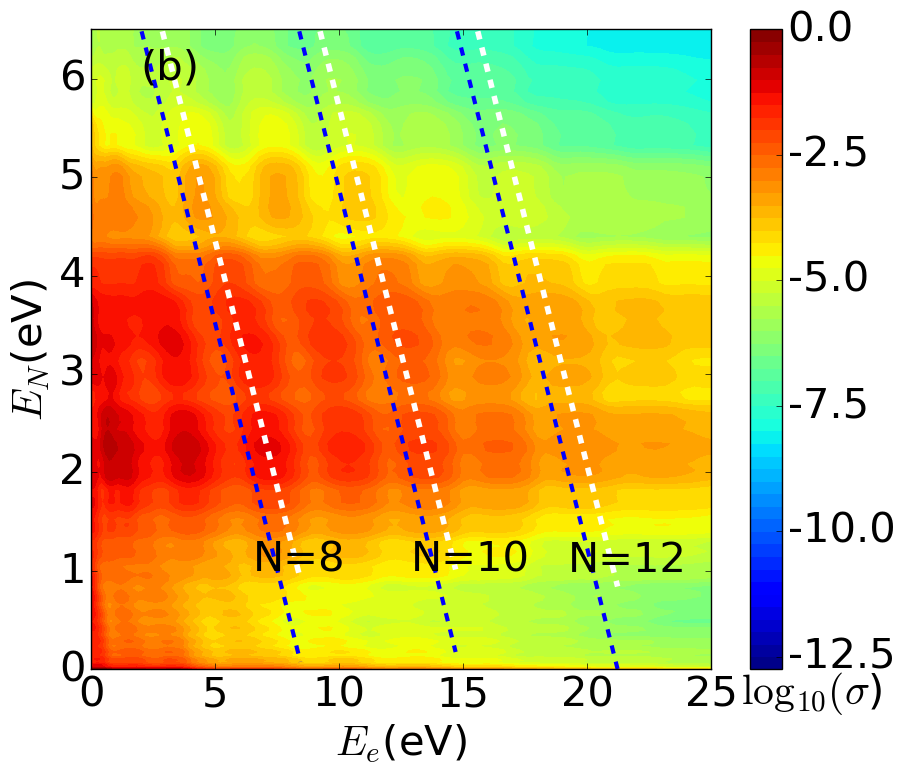}
\caption{Log-scale JES $\log_{10}\sigma(E_{N},E_{e})$ represented by total energy of two protons $E_{N}$ and that of an electron $E_{e}$.
Linear polarized, 390 nm at $I=\inten{5.9}{13}$, with an (a) $\cos^8$ envelope, with FWHM=13 $\optc$ as used in Ref.~\cite{Wu2013}, and a (b) flat-top envelope FWHM=13 $\optc$ laser pulses are applied to the $\hydroplus$ ion. The blue dashed lines represent the energy sharing between the protons and electron with formula $E_{N}+E_{e}=N\omega+E_0-U_p$ and white ones represent $E_{N}+E_{e}=N\omega+E_0$, where $\omega$ is the photon energy.
}
\label{fig:H2PlusJESExp}
\end{figure}
}
\subsection{wavefunction analysis}
The resonance is usually accompanied with states mixing which contributes to populations of excited states that survives after the pulse.
The probability distributions of the electron and the protons are calculated by integrating the 6D wavefunction on radial coordinates $r_{1/2}\in[R_0,R_1]$ and the whole angular coordinates as
\begin{equation}\label{eq:probabilityN}
\begin{split}
 p_{N}(\phi_1,\theta_1, r_1, R_0,R_1, T)=&\int d\phi_2 \int \sin\theta_2 d\theta_2\\
 &\int_{R_0}^{R_1} r_2^2 dr_2|\psi(\vec{r_1},\vec{r_2},t)|^2
 \end{split}
\end{equation}
for the protons, and
\begin{equation}\label{eq:probabilitye}
 \begin{split}
 p_{e}(\phi_2,\theta_2, r_2, R_0,R_1, T)=&\int d\phi_1 \int \sin\theta_1 d\theta_1 \\
 &\int_{R_0}^{R_1} r_1^2 dr_1|\psi(\vec{r_1},\vec{r_2},t)|^2
\end{split}
\end{equation}
for the electron.
We split the radial coordinates into the inner region and the outer region $r_{1,2}\in[R_0,R_1]$, and the yields of both regions are normalized by diving the maximum probability of the region over the flat-top and $\cos^8$ envelopes.
We only focus on the wave packets after the time-propagation at $t=T$ ($T\geq 330$ atomic units for flat-top and $T\geq770$ atomic units for $\cos^8$ envelopes), because they are important for investigating the resonance.
\par
\begin{figure} 
\centering
\includegraphics[width=0.4\textwidth,trim=0.1cm 0.1cm 0.1cm 0.1cm,clip]{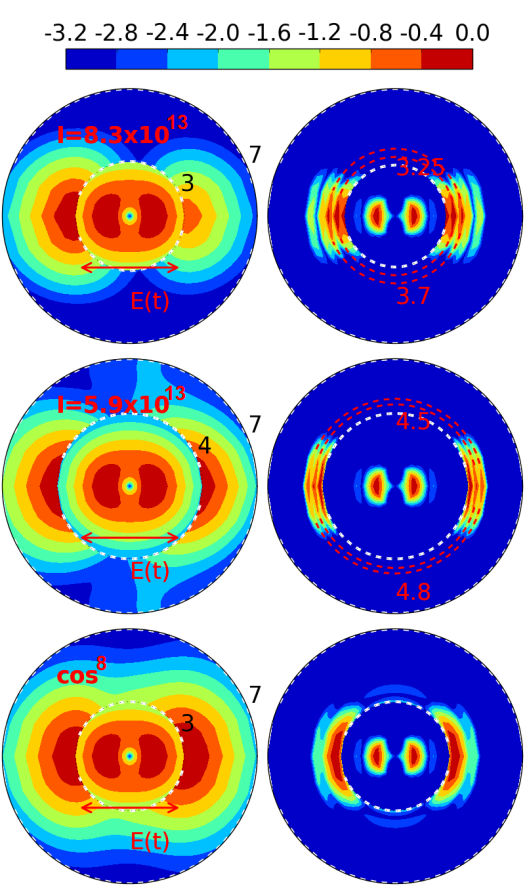}
\caption{The log-scale probability distribution of electron by Eq.~\ref{eq:probabilitye} (left column) and of protons by Eq.~\ref{eq:probabilityN} (right column) and with flat-top envelope at $\inten{8.3}{13}$ (first row), $\inten{5.9}{13}$ (second row) and $\cos^8$ envelope at $\inten{8.3}{13}$ (third row).
$[R_0,R_1]$ are $[0,3]$ and $[3,7]$ atomic units for the inner and outer regions of the first and third rows; $[R_0,R_1]$ are $[0,4]$ and $[4,7]$ atomic units for inner and outer regions of the second row.
The inner and outer regions are split by the white, dashed circle along with the radial values, which are depicted in the electron figure in the left column; values of each region are both normalized by dividing the maximum number.
The radial positions enhanced yields of the protons in the outside region by Eq.~\ref{eq:probabilityN} are illustrated by the dashed red circles and ticks in the right column.
The absolute value of the outer shell is several orders smaller than the inner shell.
The peaks of yields of the protons near the white circle are neglected because it comes from the normalization.
The polarization direction is along the horizontal axis and the direction electric field is labeled at each sub-figure with an arrow above "$E(t)$".
}
\label{fig:wavefunctionEvolution}
\end{figure}

The wavefunction evolution of the electron and the protons at the end of the pulse is illustrated in Fig.~\ref{fig:wavefunctionEvolution}.
Before analyzing the figure, we would like to point out that the values of the two regions are normalized in order to make the illustration better.
The absolute values in the outer region are insignificant compared to those of the inner region; the absolute values of the outer region of the third row are much smaller than those of the first two rows.
For the distribution of the electron after the pulse in the left column of Fig.~\ref{fig:wavefunctionEvolution}, the yields of the inner region are similar for the flat-top and the $\cos^8$ envelopes; in the outer region, the distributions are less symmetric for the flat-top envelope, indicating a dominance for ungerade $^2\Sigma_u^+$ states.
For the distribution of protons after the pulse, we scan the $R_0,R_1$ values and only find the existence of enhanced yields at two radial values as depicted in the right column of Fig.~\ref{fig:wavefunctionEvolution}.
There exist discrete peaks of yields for fat-top envelop computation, with $r_1=3.25,3.7$ ($R=2r_1=6.5,7.4$) for $I=\inten{8.3}{13}$, and $r_1=4.5,4.8$ ($R=2r_1=9,9.6$) for $I=\inten{5.9}{13}$ atomic units.
\rev{These enhanced yields could serve as evidences for states mixing and Freeman resonances.}
However, for the $\cos^8$ envelope computation, the yields are contiguous \rev{and no such enhanced yields are observed}.
\par
\begin{figure} 
\centering
\includegraphics[width=0.45\textwidth]{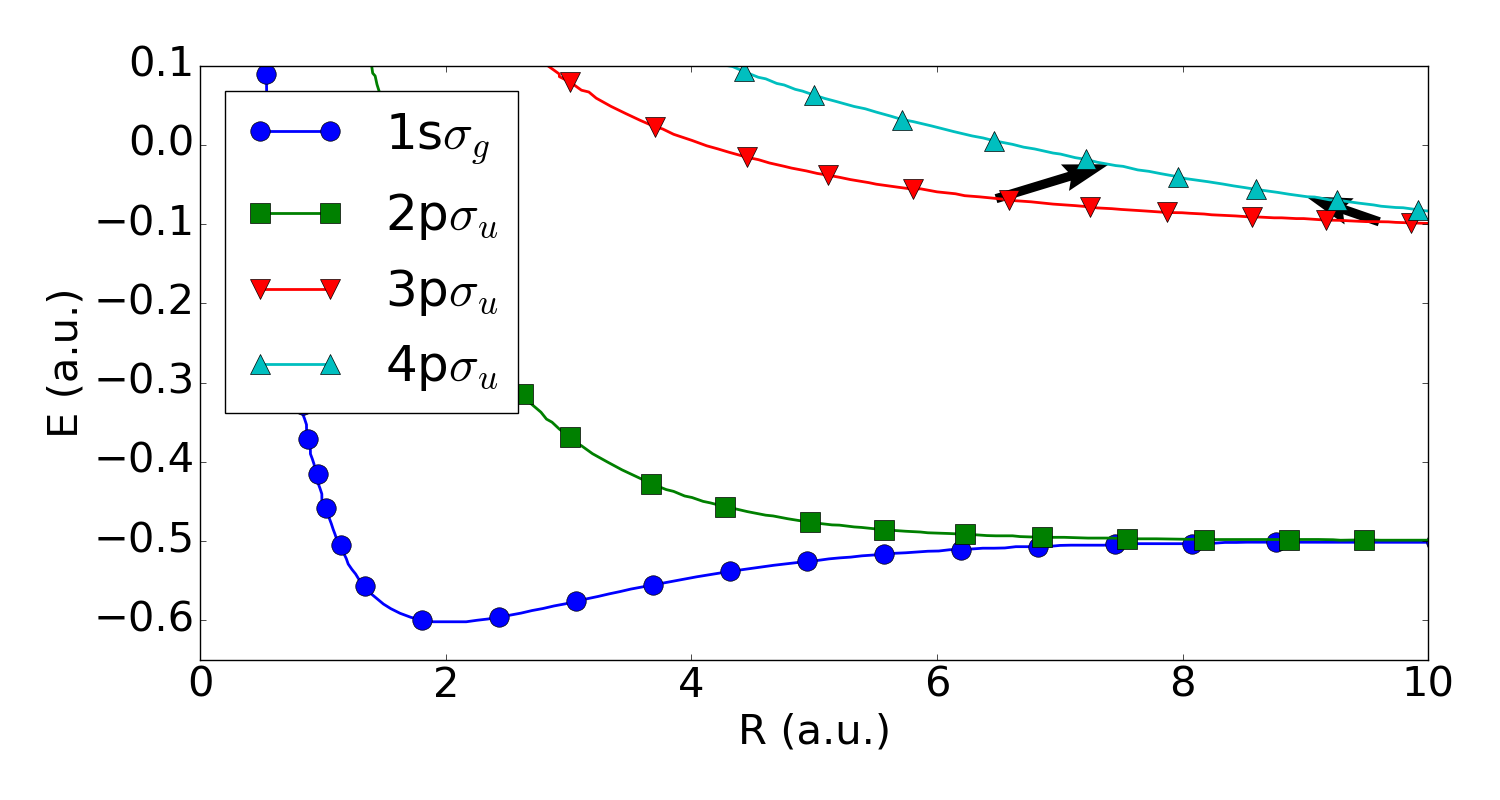}
\caption{The eigenstates of electrons of $\hydroplus$ with internuclear distance R \rev{are calculated from the generalized Lambert W function~\cite{Scott2006}}. The two vectors depict the energy transfer with the help of ponderomotive energy $U_p$. The energy transfer $4p\sigma_u(7.4)-3p\sigma_u(6.5)=0.0434$ atomic units with $U_p(I=\inten{8.3}{13})=0.0455$ atomic units
and $4p\sigma_u(9.6)-3p\sigma_u(9)=0.0322$ atomic units with $U_p(I=\inten{5.9}{13})=0.0330$ atomic units. The small error may come from the neglect of $E_N$, which is not included in this figure.
}
\label{fig:eigens}
\end{figure}

\rev{Considering the radial position of the peaks of the enhanced yields,}
we find that the relative energy of the two ungerade states at the enhanced radial coordinates above are the ponderomotive energies, as is depicted by the two black vectors in Fig.~\ref{fig:eigens}.
\rev{Thus the enhanced yields in the analysis of wavefunction is highly correlated with the energy transfer between these two excited states, and $U_p$ is consumed during this process.
This contributes to the "missing of $U_p$" in JES.}
The internuclear distance of protons are changed during the resonance and more intermediate, excited states of protons are created.
This \rev{also} explains the higher yield in JES with nuclear KER $\geq 3$ eV for the flat-top envelope than the $\cos^8$ envelope in Fig.~\ref{fig:H2PlusJES}.
\par
\rev{
To sum up, from the analysis of the wavefunction after the end of the pulse, we find a clear signature of Freeman resonance in $\hydroplus$ when shot by the flat-top envelope laser pulse, compared to the short, $\cos^8$ envelope laser pulse.
The energy transfer between the two corresponding excited states contributes to the "missing of $U_p$" in the JES.
}

\section{Conclusion and discussion}
We computed the JES for a 400 nm pulse with flat-top envelope and $\cos^8$ envelope at $\inten{8.3}{13}$ and $\inten{5.9}{13}$.
In JES, the energy sharing of $N$ photons with frequency $\omega$ by nuclear KER $E_N$ and electronic KER $E_e$ are well represented by $E_N+E_e=N\omega+E_0-U_p$ for $\cos^8$ pulses, but satisfy $E_N+E_e=N\omega+E_0$ for flat-top envelope.
The difference comes from the ungerade excited states with energy difference $U_p$.
We propose that this is a universal effect for an arbitrary intensity as the eigenenergies of $\hydroplus$ are contiguous with the bound length $R$.
At higher intensities, there may also exist other resonance states with energy difference $N\omega+U_p$, which requires further investigation. 
\section*{Acknowledgments}
J.Z. was supported by the DFG Priority Programme 1840, QUTIF.
We are grateful for fruitful discussions with Dr. Lun Yue from Louisiana State University, Dr. Xiaochun Gong and Dr. Hongcheng Ni from East China Normal University, and Prof. Dr. Armin Scrinzi from Ludwig Maximilians University.

\bibliography{h2plus.bib}
\end{document}